\documentclass[11pt,a4paper]{article}
\usepackage{float}
\usepackage{epsfig}
\usepackage{multirow}
\usepackage{multicol}
\usepackage{graphics}
\usepackage[cp1252,utf8]{inputenc}
\usepackage{hyperref}
\usepackage{graphicx}
\usepackage{amsmath}
\usepackage{amsfonts}
\usepackage{epstopdf}
\usepackage{xcolor}
\usepackage{amssymb}
\usepackage{bigints} 
\usepackage{relsize} 
\usepackage[top=1in, bottom=2cm, left=1.2cm, right=1.2cm]{geometry}
\usepackage{authblk}
\usepackage{cite}

\usepackage{changepage} 
\usepackage{array, makecell} %

\usepackage[toc,page]{appendix}
\usepackage{afterpage}

\usepackage{placeins}
\usepackage{caption}
\usepackage{wrapfig}
\usepackage{subcaption}
\usepackage{soul}
\usepackage{cancel}
\usepackage{ulem}
\usepackage[utf8]{inputenc}
\usepackage[english]{babel}

\usepackage{xcolor}
\newcommand{\newc}{\newcommand}

\def\Ord{\lower .7ex\hbox{$\;\stackrel{\textstyle <}{\sim}\;$}}
\def\OOrd{\lower .7ex\hbox{$\;\stackrel{\textstyle >}{\sim}\;$}}

\newc{\order}{{\cal O}}

\newc{\be}{\begin{equation}}
\newc{\ee}{\end{equation}}
\newc{\br}{\begin{eqnarray}}
\newc{\er}{\end{eqnarray}}
\newc{\ba}{\begin{array}}
\newc{\ea}{\end{array}}
\newc{\bi}{\begin{itemize}}
\newc{\ei}{\end{itemize}}
\newc{\bn}{\begin{enumerate}}
\newc{\en}{\end{enumerate}}
\newc{\bc}{\begin{center}}
\newc{\ec}{\end{center}}
\newc{\ra}{\rightarrow}
\newc{\lra}{\longrightarrow}
\newc{\wt}{\widetilde}
\newc{\til}{\tilde}

\newc{\wh}{\widehat}
\newc{\ti}{\times}
\newc{\Dir}{\kern -6.4pt\Big{/}}
\newc{\Dirin}{\kern -10.4pt\Big{/}\kern 4.4pt}
\newc{\DDir}{\kern -10.6pt\Big{/}}
\newc{\DGir}{\kern -6.0pt\Big{/}}
\newc{\sig}{\sigma}
\newc{\sigmalstop}{\sig_{\lstoppair}}
\newc{\Sig}{\Sigma}  
\newc{\del}{\delta}
\newc{\Del}{\Delta}
\newc{\lam}{\lambda}
\newc{\Lam}{\Lambda}
\newc{\gam}{\gamma}
\newc{\Gam}{\Gamma}
\newc{\eps}{\epsilon}
\newc{\Eps}{\Epsilon}
\newc{\kap}{\kappa}
\newc{\Kap}{\Kappa}
\newc{\modulus}[1]{\left| #1 \right|}
\newc{\eq}[1]{(\ref{eq:#1})}
\newc{\eqs}[2]{(\ref{eq:#1},\ref{eq:#2})}
\newc{\etal}{{\it et al.}\ }
\newc{\ibid}{{\it ibid}.}
\newc{\ibidem}{{\it ibidem}.}
\newc{\eg}{{\it e.g.}\ }
\newc{\ie}{{\it i.e.}\ }

\newc{\nonum}{\nonumber}
\newc{\lab}[1]{\label{eq:#1}}
\newc{\dpr}[2]{({#1}\cdot{#2})}
\newc{\lt}{\stackrel{<}}
\newc{\gt}{\stackrel{>}}
\newc{\lsimeq}{\stackrel{<}{\sim}}
\newc{\gsimeq}{\stackrel{>}{\sim}}
\def\lsim{\buildrel{\scriptscriptstyle <}\over{\scriptscriptstyle\sim}}
\def\gsim{\buildrel{\scriptscriptstyle >}\over{\scriptscriptstyle\sim}}
\def\lapp{\mathrel{\rlap{\raise.5ex\hbox{$<$}}
                    {\lower.5ex\hbox{$\sim$}}}}
\def\gapp{\mathrel{\rlap{\raise.5ex\hbox{$>$}}
                    {\lower.5ex\hbox{$\sim$}}}}
\newc{\half}{\frac{1}{2}}

\newc{\bQ}{\ol{Q}}
\newc{\dota}{\dot{\alpha }}
\newc{\dotb}{\dot{\beta }}
\newc{\dotd}{\dot{\delta }}
\newc{\nindnt}{\noindent}

\newc{\matth}{\mathsurround=0pt}
\def\ML{\ifmmode{{\mathaccent"7E M}_L}
             \else{${\mathaccent"7E M}_L$}\fi}
\def\MR{\ifmmode{{\mathaccent"7E M}_R}
             \else{${\mathaccent"7E M}_R$}\fi}

\newc{\mr}{\mathrm}

\newc{\siminf}{\mbox{$_{\sim}$ {\small {\hspace{-1.em}{$<$}}}    }}
\newc{\simsup}{\mbox{$_{\sim}$ {\small {\hspace{-1.em}{$>$}}}    }}

\newc {\Zboson}{{\mathrm Z}^{0}}
\newc{\thetaw}{\theta_W}
\newc{\mbot}{{m_b}}
\newc{\mtop}{{m_t}}
\newc{\sm}{${\cal {SM}}$}
\newc{\as}{\alpha_s}
\newc{\aem}{\alpha_{em}}

\newc{\ppbar}{\mbox{$p\ol{p}$}}
\newc{\bbbar}{\mbox{$b\ol{b}$}}
\newc{\ccbar}{\mbox{$c\ol{c}$}}
\newc{\ttbar}{\mbox{$t\ol{t}$}}
\newc{\eebar}{\mbox{$e\ol{e}$}}
\newc{\zzero}{\mbox{$Z^0$}}

\newc{\wplus}{\mbox{$W^+$}}
\newc{\wminus}{\mbox{$W^-$}}
\newc{\ellp}{\ell^+}
\newc{\ellm}{\ell^-}
\newc{\elp}{\mbox{$e^+$}}
\newc{\elm}{\mbox{$e^-$}}
\newc{\elpm}{\mbox{$e^{\pm}$}}
\newc{\qbar}     {\mbox{$\ol{q}$}}

\newc{\Ebar}{{\bar E}}
\newc{\Dbar}{{\bar D}}
\newc{\Ubar}{{\bar U}}
\newc{\susy}{{{SUSY}}}
\newc{\msusy}{{{M_{SUSY}}}}

\def\photino{\ifmmode{\mathaccent"7E \gam}\else{$\mathaccent"7E \gam$}\fi}
\def\taugluino{\ifmmode{\tau_{\mathaccent"7E g}}
             \else{$\tau_{\mathaccent"7E g}$}\fi}
\def\mphotino{\ifmmode{m_{\mathaccent"7E \gam}}
             \else{$m_{\mathaccent"7E \gam}$}\fi}
\newc{\gl}   {\mbox{$\wt{g}$}}
\newc{\mgl}  {\mbox{$m_{\gl}$}}

\def \chonep {{\wt\chi_1^+}}

\def \ch2p {{\wt\chi_2^+}}
\def \chonem {{\wt\chi_1^-}}
\def \ch2m {{\wt\chi_2^-}}

\def \chonepm{{\wt\chi_1}^{\pm}}

\newc{\dmchi}{\Delta m_{\wt\chi}}

\def \lspone{\wt\chi_1^0}
\def \mlspone{m_{\lspone}}

\newc{\sele}{\wt{\mathrm e}}
\newc{\sell}{\wt{\ell}}

\newc{\snue}     {\mbox{$ \wt{\nu_e}$}}
\newc{\smu}{\wt{\mu}}
\newc{\stau}{\wt{\tau}}
\newc {\nuL} {\wt{\nu}_L}
\newc {\nuR} {\wt{\nu}_R}
\newc {\snub} {\bar{\wt{\nu}}}
\newc {\eL} {\wt{e}_L}
\newc {\eR} {\wt{e}_R}

\def \stau{\wt\tau}

\def \sq{\wt{q}}

\newc{\msqot}  {\mbox{$m_(\sq_{1,2} )$}}
\newc{\sqbar}    {\mbox{$\bar{\wt{q}}$}}
\newc{\ssb}      {\mbox{$\squark\ol{\squark}$}}
\newc {\qL} {\wt{q}_L}
\newc {\qR} {\wt{q}_R}
\newc {\uL} {\wt{u}_L}
\newc {\uR} {\wt{u}_R}

\newc {\dL} {\wt{d}_L}
\newc {\dR} {\wt{d}_R}
\newc {\cL} {\wt{c}_L}
\newc {\cR} {\wt{c}_R}
\newc {\sL} {\wt{s}_L}
\newc {\sR} {\wt{s}_R}
\newc {\tL} {\wt{t}_L}
\newc {\tR} {\wt{t}_R}
\newc {\stb} {\ol{\wt{t}}_1}
\newc {\sbot} {\wt{b}_1}
\newc {\msbot} {m_{\sbot}}
\newc {\sbotb} {\ol{\wt{b}}_1}
\newc {\bL} {\wt{b}_L}
\newc {\bR} {\wt{b}_R}

\newc{\csquark}  {\mbox{$\wt{c}$}}
\newc{\csquarkl} {\mbox{$\wt{c}_L$}}
\newc{\mcsl}     {\mbox{$m(\csquarkl)$}}

\newc {\stopl}         {\wt{\mathrm{t}}_{\mathrm L}}
\newc {\stopr}         {\wt{\mathrm{t}}_{\mathrm R}}
\newc {\stoppair}      {\wt{\mathrm{t}}_{1}
\bar{\wt{\mathrm{t}}}_{1}}
\def \lstop{\wt{t}_{1}}

\def \lstoppair{\lstop\lstop^*}

\newc{\tsquark}  {\mbox{$\wt{t}$}}
\newc{\ttb}      {\mbox{$\tsquark\ol{\tsquark}$}}
\newc{\ttbone}   {\mbox{$\tsquark_1\ol{\tsquark}_1$}}

\newc{\mix}{\theta_{\wt t}}
\newc{\cost}{\cos{\theta_{\wt t}}}
\newc{\sint}{\sin{\theta_{\wt t}}}
\newc{\costloop}{\cos{\theta_{\wt t_{loop}}}}

\newc{\mixsbot}{\theta_{\wt b}}

\newc{\tb}{\tan\beta}
\newc{\cb}{\cot\beta}
\newc{\vev}[1]{{\left\langle #1\right\rangle}}

\newc{\mhalf}{m_{1/2}}
\newc{\mzero} {\mbox{$m_0$}}
\newc{\azero} {\mbox{$A_0$}}

\newc{\lb}{\lam}
\newc{\lbp}{\lam^{\prime}}
\newc{\lbpp}{\lam^{\prime\prime}}
\newc{\rpv}{{\not \!\! R_p}}
\newc{\rpvm}{{\not  R_p}}
\newc{\rp}{R_{p}}
\newc{\rpmssm}{{RPC MSSM}}
\newc{\rpvmssm}{{RPV MSSM}}

\newc{\sbyb}{S/$\sqrt B$}
\newc{\pelp}{\mbox{$e^+$}}
\newc{\pelm}{\mbox{$e^-$}}
\newc{\pelpm}{\mbox{$e^{\pm}$}}
\newc{\epem}{\mbox{$e^+e^-$}}
\newc{\lplm}{\mbox{$\ell^+\ell^-$}}

\def\Ecm{\ifmmode{E_{\mathrm{cm}}}\else{$E_{\mathrm{cm}}$}\fi}
\newc{\rts}{\sqrt{s}}
\newc{\rtshat}{\sqrt{\hat s}}
\newc{\gev}{\,GeV}
\newc{\mev}{~{\rm MeV}}
\newc{\tev}  {\mbox{$\;{\rm TeV}$}}
\newc{\gevc} {\mbox{$\;{\rm GeV}/c$}}
\newc{\gevcc}{\mbox{$\;{\rm GeV}/c^2$}}
\newc{\intlum}{\mbox{${ \int {\cal L} \; dt}$}}
\newc{\call}{{\cal L}}
\def \met  {\mbox{${E\!\!\!\!/_T}$}}

\newc{\ptmiss}{/ \hskip-7pt p_T}

\newc{\PT}{\mbox{$p_T$}}
\newc{\ET}{\mbox{$E_T$}}
\newc{\dedx}{\mbox{${\rm d}E/{\rm d}x$}}
\newc{\ifb}{\mbox{${\rm fb}^{-1}$}}
\newc{\ipb}{\mbox{${\rm pb}^{-1}$}}
\newc{\pb}{~{\rm pb}}
\newc{\fb}{~{\rm fb}}
\newc{\ycut}{y_{\mathrm{cut}}}
\newc{\chis}{\mbox{$\chi^{2}$}}

\def \jet(s){\emph{jet(s) }}

\newc{\mpl}{M_{\rm Pl}}
\newc{\mgut}{M_{GUT}}
\newc{\mw}{M_{W}}
\newc{\mweak}{M_{weak}}
\newc{\mz}{M_{Z}}

\newc{\OPALColl}   {OPAL Collaboration}
\newc{\ALEPHColl}  {ALEPH Collaboration}
\newc{\DELPHIColl} {DELPHI Collaboration}
\newc{\XLColl}     {L3 Collaboration}
\newc{\JADEColl}   {JADE Collaboration}
\newc{\CDFColl}    {CDF Collaboration}
\newc{\DXColl}     {D0 Collaboration}
\newc{\HXColl}     {H1 Collaboration}
\newc{\ZEUSColl}   {ZEUS Collaboration}
\newc{\LEPColl}    {LEP Collaboration}
\newc{\ATLASColl}  {ATLAS Collaboration}
\newc{\CMSColl}    {CMS Collaboration}
\newc{\UAColl}    {UA Collaboration}
\newc{\KAMLANDColl}{KamLAND Collaboration}
\newc{\IMBColl}    {IMB Collaboration}
\newc{\KAMIOColl}  {Kamiokande Collaboration}
\newc{\SKAMIOColl} {Super-Kamiokande Collaboration}
\newc{\SUDANTColl} {Soudan-2 Collaboration}
\newc{\MACROColl}  {MACRO Collaboration}
\newc{\GALLEXColl} {GALLEX Collaboration}
\newc{\GNOColl}    {GNO Collaboration}
\newc{\SAGEColl}  {SAGE Collaboration}
\newc{\SNOColl}  {SNO Collaboration}
\newc{\CHOOZColl}  {CHOOZ Collaboration}
\newc{\PDGColl}  {Particle Data Group Collaboration}

\def\issue(#1,#2,#3){{\bf #1}, #2 (#3)}
\def\iss(#1,#2,#3){{\bf #1} (#3) #2}
\def\ASTR(#1,#2,#3){Astropart.\ Phys. \issue(#1,#2,#3)}
\def\AJ(#1,#2,#3){Astrophysical.\ Jour. \issue(#1,#2,#3)}
\def\AJS(#1,#2,#3){Astrophys.\ J.\ Suppl. \issue(#1,#2,#3)}
\def\APP(#1,#2,#3){Acta.\ Phys.\ Pol. \issue(#1,#2,#3)}
\def\JCAP(#1,#2,#3){Journal\ XX. \issue(#1,#2,#3)} 
\def\SC(#1,#2,#3){Science \issue(#1,#2,#3)}
\def\PRD(#1,#2,#3){Phys.\ Rev.\ D \issue(#1,#2,#3)}
\def\PR(#1,#2,#3){Phys.\ Rev.\ \issue(#1,#2,#3)} 
\def\PRC(#1,#2,#3){Phys.\ Rev.\ C \issue(#1,#2,#3)}
\def\NPB(#1,#2,#3){Nucl.\ Phys.\ B \issue(#1,#2,#3)}
\def\NPPS(#1,#2,#3){Nucl.\ Phys.\ Proc. \ Suppl \issue(#1,#2,#3)}
\def\NJP(#1,#2,#3){New.\ J.\ Phys. \issue(#1,#2,#3)}
\def\JP(#1,#2,#3){J.\ Phys.\issue(#1,#2,#3)}
\def\PL(#1,#2,#3){Phys.\ Lett. \issue(#1,#2,#3)}
\def\ZP(#1,#2,#3){Z.\ Phys. \issue(#1,#2,#3)}
\def\ZPC(#1,#2,#3){Z.\ Phys.\ C  \issue(#1,#2,#3)}
\def\PREP(#1,#2,#3){Phys.\ Rep. \issue(#1,#2,#3)}
\def\PRL(#1,#2,#3){Phys.\ Rev.\ Lett. \issue(#1,#2,#3)}
\def\MPL(#1,#2,#3){Mod.\ Phys.\ Lett. \issue(#1,#2,#3)}
\def\RMP(#1,#2,#3){Rev.\ Mod.\ Phys. \issue(#1,#2,#3)}
\def\SJNP(#1,#2,#3){Sov.\ J.\ Nucl.\ Phys. \issue(#1,#2,#3)}
\def\CPC(#1,#2,#3){Comp.\ Phys.\ Comm. \issue(#1,#2,#3)}
\def\IJMPA(#1,#2,#3){Int.\ J.\ Mod. \ Phys.\ A \issue(#1,#2,#3)}
\def\MPLA(#1,#2,#3){Mod.\ Phys.\ Lett.\ A \issue(#1,#2,#3)}
\def\PTP(#1,#2,#3){Prog.\ Theor.\ Phys. \issue(#1,#2,#3)}
\def\RMP(#1,#2,#3){Rev.\ Mod.\ Phys. \issue(#1,#2,#3)}
\def\NIMA(#1,#2,#3){Nucl.\ Instrum.\ Methods \ A \issue(#1,#2,#3)}
\def\EPJC(#1,#2,#3){Eur.\ Phys.\ J.\ C \issue(#1,#2,#3)}
\def\RPP (#1,#2,#3){Rept.\ Prog.\ Phys. \issue(#1,#2,#3)}
\def\PPNP(#1,#2,#3){ Prog.\ Part.\ Nucl.\ Phys. \issue(#1,#2,#3)}
\newc{\PRDR}[3]{{Phys. Rev. D} {\bf #1}, Rapid  Communications, #2 (#3)}

\def\PLB(#1,#2,#3){Phys.\ Lett.\ B  \iss(#1,#2,#3)}
\def\JHEP(#1,#2,#3){JHEP \iss(#1,#2,#3)}

\def\gmin2{(g-2)_\mu}

\catcode`\@=11 

\def\vev#1{\left\langle #1\right\rangle}
\def\lsim{\mathrel{\mathpalette\@versim<}}
\def\gsim{\mathrel{\mathpalette\@versim>}}
\def\@versim#1#2{\vcenter{\offinterlineskip
    \ialign{$\m@th#1\hfil##\hfil$\crcr#2\crcr\sim\crcr } }}
\def\etal{{\em et. al.}}

\def\r2{\sqrt 2}
\def\beq{\begin{equation}}
\def\eeq{\end{equation}}
\def\beqn{\begin{eqnarray}}
\def\eeqn{\end{eqnarray}}
\def\sinW2{\sin^2\theta_W}

\def\mz2{M_{z}^2}
\def\c2b{\cos 2\beta}

\def\m#1{{\tilde m}_#1}

\def\mw#1{{\tilde m}_{\omega #1}}

\def\mz{M_Z}
\def\m0{m_0}
\def\mhalf{m_{\frac{1}{2}}}

\def\cb{\cos\beta}

\def\sec2w{sec^2\theta_W}

\def\gmin2{(g-2)_\mu}

\catcode`\@=11 

\def\vev#1{\left\langle #1\right\rangle}
\def\lsim{\mathrel{\mathpalette\@versim<}}
\def\gsim{\mathrel{\mathpalette\@versim>}}
\def\@versim#1#2{\vcenter{\offinterlineskip
    \ialign{$\m@th#1\hfil##\hfil$\crcr#2\crcr\sim\crcr } }}
\def\etal{{\em et. al.}}

\def\tb{\tilde b}

\def\tL{\tilde L}

\def \chonep{{\wt\chi_1}^{+}}
\def \chonem{{\wt\chi_1^-}}
\def \chonep2{{\wt\chi_2^+}}
\def \chonem2{{\wt\chi_2^-}}

\def \chonepm{{\wt\chi_1}^{\pm}}

\def \lstop{\wt{t}_{1}}

\def \lspone{\wt\chi_1^0}
\def \mlspone{m_{\lspone}}

\def\PL{Phys. Lett.}
\def\PRL{Phys. Rev. Lett.}

\def\PR{Phys. Rev.}

\def \lspone{\wt\chi_1^0}
\def \chonem {{\wt\chi_1^\pm}}
\def \chargino1 {{\wt\chi_1^\pm}}
\def \chargino2 {{\wt\chi_2^\pm}}
\def \lstop{\wt{t}_{1}}
\def \ch2m {{\wt\chi_2^-}}

\def \chonep {{\wt\chi_1^+}}

\providecommand{\keywords}[1]
{
  \small	
  \textbf{\textit{Keywords---}} #1
}

	\title{\bf Revisiting the  gluino mass 
		limits in the pMSSM in the light of the latest LHC data and Dark Matter constraints }

	\author[1]{\bf Abhi Mukherjee \thanks{abhiphys18@klyuniv.ac.in}}
	\author[2]{\bf Saurabh Niyogi \thanks{saurabhphys@gmail.com}}
	\author[3]{\bf Sujoy Poddar \thanks{sujoy.phy@gmail.com}}
	\affil[1]{Department of Physics, University of Kalyani, Kalyani 741235, India}
	\affil[2]{Gokhale Memorial Girls' College, 1/1 Harish Mukherjee Road, Kolkata 700 020, India}
	\affil[3]{Department of Physics, Diamond Harbour Women's University, Diamond Harbour Road, Sarisha, South 24 Parganas, West Bengal 743368, India}

	\date{}
 
\begin{document}

	\maketitle
	\begin{abstract}
	\noindent The purpose of this paper is to examine the model dependence of the stringent constraints on the gluino mass obtained from the Large Hadron Collider (LHC) experiments by analyzing the Run II  data using specific simplified models based on several ad hoc sparticle spectra which cannot be realized even in the fairly generic pMSSM models. We first revisit the bounds on the gluino mass placed by the ATLAS collaboration using the $1l + jets + \met$ data. We show that the exclusion region in the $M_{\widetilde{g}}-M_{\widetilde{\chi}^0_1}$ plane in the pMSSM scenario sensitively depends on the mass hierarchy between the  left and right squarks and composition of the lighter electroweakinos and, to a lesser extent, other parameters. Most importantly, for higgsino type lighter electroweakinos (except for the LSP), the bound on the gluino mass from this channel practically disappears. However, if such models are confronted by  the ATLAS $jets + \met$ data, fairly strong limits are regained. Thus in the pMSSM an analysis involving a small number of channels may provide more reliable mass limits. We have also performed detailed analyses on neutralino dark matter (DM) constraints in the models we have studied and have found that for a significant range of LSP masses, the relic density constraints from  the WMAP/PLANCK data are satisfied and LSP-gluino coannihilation plays an important role in relic density production. We have also checked the simultaneous compatibility of the models studied here with the direct DM detection, and the LHC constraints.
	\end{abstract}


	\section{Introduction}

Supersymmetry (SUSY)\cite{Nilles:1983ge,Haber:1984rc,bagger1992theory,Lykken:1996xt,Martin:1997ns,Chung:2003fi,drees2005theory,Baer:2006rs}, which allows inter-conversion of fermions and bosons, is one of the most well motivated and widely studied extension of the Standard Model (SM). It offers solutions to several short-comings of the SM like, the hierarchy problem\cite{Gildener:1976ih,PhysRevD.14.1667}, the unification of gauge couplings\cite{Ellis:1990wk,Amaldi:1991cn} at a scale, $M_G\simeq10^{16}$ GeV and it contains a viable Dark Matter (DM) candidate \cite{Jungman:1995df,Lahanas:2003bh,Freedman:2003ys,Roszkowski:2004jc,Bertone:2004pz,Olive:2005qz,Baer:2008uu,Drees:2012ji,Arrenberg:2013rzp}. SUSY can also trigger Electro-Weak Symmetry Breaking (EWSB)\cite{Ibanez:1982fr,Inoue:1983pp,Ibanez:1983di} which is an ad hoc phenomenon within the realm of the SM.  SUSY being the most popular and attractive extension of the SM, the ATLAS and the CMS collaborations at the Large Hadron Collider (LHC) experiments have been  trying to probe various SUSY scenarios in different channels in order to pin down the physics beyond the SM (BSM). However, so far the LHC has not been able to come up with any hint of SUSY up to center-of-mass energy of 13 TeV yet. 

Sparticles like squarks and gluinos belonging to the strong sector of SUSY have been searched with great enthusiasm  because of their large production cross sections. Consequently stringent bounds on the masses of these sparticles have been obtained by both ATLAS and CMS collaborations from RUN I\cite{TheATLAScollaboration:2013fha,TheATLAScollaboration:2013tha,TheATLAScollaboration:2013uha,ATLAS:2013tma,Aad:2013wta,ATLAS:2013ama,CMS:2013cfa} and RUN II\cite{Aad:2016eki,Aad:2016qqk,Aaboud:2016zdn,Aad:2016tuk,Aaboud:2017bac,Aaboud:2017ayj,Aaboud:2017dmy,Aaboud:2017vwy,Aaboud:2017hrg,Aaboud:2018ujj,Aaboud:2018mna,Aad:2019ftg,Aad:2020nyj,Aad:2021egl,Aad:2021jmg} data \footnote{ In a recent extensive study, the pathways of the SUSY spanning both theoretical scenarios and experimental signatures, can be found in ref. \cite{Adam:2021rrw}. Moreover, phenomenological study of the Minimal Supersymmetric Standard Model (MSSM) in future colliders can be found in ref. \cite{Abdussalam:2019pwu}}.
However, these bounds are obtained on the basis of certain assumptions tailor-made for analyzing and interpreting experimental data in an effective manner. This is termed as simplified model which is described by a smaller set of parameters in terms of masses, couplings, branching ratios and cross sections. Though simplified models provide a useful starting point for characterizing signals of new physics, often they seem somewhat ad hoc since the assumptions made in such models are hard to realize in generic scenarios, like the phenomenological Minimal Supersymmetric Standard Model (pMSSM) \cite{Djouadi:1998di}, a popular and broader framework for studying SUSY. Thus, relaxation of the bounds arising from simplified models is possible in many ways. Compressed SUSY scenarios are prime examples where such relaxation of bounds are possible \cite{LeCompte:2011fh,Dreiner:2012gx,Bhattacherjee:2012mz,Bhattacherjee:2013wna,Cohen:2013xda,Mukhopadhyay:2014dsa,Low:2014cba}.

{Recently the ATLAS collaboration has published their analyses for squark/gluino searches based on integrated luminosity of 139 $\ifb$ at the LHC RUN II which, as expected puts stronger mass bounds on squarks and gluinos. In this paper we begin with the channel $1l+jets+\cancel{E_T}$ \cite{Aad:2021zyy}. In the ref.\cite{Aad:2021zyy} the results are interpreted in terms of a simplified models by the ATLAS collaboration.} In the  ATLAS analysis the relevant sparticles are assumed to be the gluino ($\widetilde{g}$), the lighter chargino ($\widetilde{\chi}_1^{\pm}$) and the lightest neutralino ($\widetilde{\chi}_1^0$), where   the masses of $\widetilde{g}$ and $\widetilde{\chi}_1^0$ are free parameters. Only the first two generations of squarks of $L$-type (the bosonic counterpart of left handed quarks) are assumed to mediate the gluino decays, while the $R$-type squarks are decoupled. In addition
$\widetilde{\chi}_1^{\pm}$ mass is assumed to be the arithmetic mean of the two free parameters. This optimizes the phase space of gluino decays. It is further assumed that the lightest supersymmetric particle (LSP) is bino like and the lighter charginos are wino like.  All other sparticle masses are set beyond the reach of the LHC. It is also assumed that
the gluinos decay into the wino like lighter charginos in association with two light quarks with  $100\%$ branching ratio (BR). The lighter charginos subsequently decay into $W^{\pm} \widetilde{\chi}_1^0$ with $100\%$ BR.  The final state lepton arises from the decay of the $W^{\pm}$ according to the SM.

It is often true that many of the above assumptions result in very strong bounds on the gluino mass. 
Some other popular frameworks, like minimal universal extra dimension (MUED) can even be ruled out based on such bounds \cite{Avnish:2020atn,Ashanujjaman:2021txz,Flores:2021xwx}.
However, if one does careful consideration of these bounds, one must, at first, recast the results of such analysis for his choice of specific models. That is the goal of this work. We will show that in a more generic and less contrived models, like pMSSM \cite{Djouadi:1998di} the above limits weaken significantly. In some cases they may even disappear. 
Secondly, although one can take simplified models as a good starting point for recasting various mass bounds at the LHC, but they are not sufficiently illuminating for studying other low energy observables, like the DM relic density constraints. As simplified models work with a few number of mass parameters, they often do not take care of the contributions from various coannihilation channels for the estimation of relic. For example, the three mass parameters relevant for the ATLAS analysis are $M_{\widetilde{g}}$, $M_{\widetilde{\chi}_1^{\pm}}$ and $M_{\widetilde{\chi}_1^{0}}$. But such parameterisation indeed misses the possible coannihilation between the DM i.e. $\widetilde{\chi}_1^{0}$ and the second lightest neutralino ($\widetilde{\chi}_2^{0}$), which is mass degenerate with the lighter chargino in pMSSM.
Hence, results in terms of simplified models are always subject to further investigations.

The purpose of this paper is to explore the above issues within the broader framework of the pMSSM. First we examine whether the above limits on the gluino mass can be significantly relaxed in some pMSSM scenarios. To begin with the pMSSM parameters are so chosen that the models resemble  some features of the ATLAS simplified model. However, some differences are inevitable. For example, if the lighter chargino is wino like, the $\widetilde{\chi}_2^0$ is also wino like and almost degenerate with the former and contributes significantly to gluino decays which affects the signal. The special choice of only relatively light $L$-type squarks, as in the ATLAS work, allows the gluino to dominantly decay into wino like sparticles. In addition the presence of relatively light $R$-type squarks enhances the fraction of direct gluino decays into the bino like LSP resulting in a further suppression of the leptonic signal. In summary in the pMSSM, gluinos may decay into all lighter electroweakinos ($\widetilde{\chi}_1^\pm/ \widetilde{\chi}_1^0/ \widetilde{\chi}_2^0$) but not with 100 \% BR in any of the three decay modes. 
Phase space suppression due to modified gluino-chargino-LSP mass hierarchies may further reduce the mass limits which we will also explore. 

Finally, the impact of changing the compositions of the lighter charginos and neutralinos (together called lighter electroweakinos) also affects the mass limits. In fact, higgsino like charginos have not received the due attention in sparticle searches at the LHC. Phenomenological studies\cite{Chakraborti:2014gea,Chakraborti:2015mra,Datta:2016ypd,Chakraborti:2017vxz,Datta:2018lup,Adam:2021rrw} for probing Electro-Weak (EW) sectors through  electroweakinos have also been performed in the light of ATLAS/CMS data\cite{Aad:2014vma,Aad:2014nua,Aad:2014yka,Aad:2015jqa}. Since the bounds on the gluino mass obtained by ATLAS is based on wino like lighter charginos, it would be interesting to see the results corresponding to a higgsino like $\widetilde{\chi}_1^{\pm}$. 

In this way we identify different classes of models where the gluino mass limits are significantly reduced. In the most dramatic cases we find that in some generic models the bounds from the $1l+jets+\met$ channels are completely washed out.

However, the class of models with suppressed $1l+jets+\met$ signal are expected to yield more $jets + \met$ events. We, therefore, revisit these models with reduced gluino mass limits using $jets + \met$ data\cite{Aad:2020aze} obtained by the ATLAS collaboration. In many cases stringent exclusion contours in the $M_{\widetilde{g}}-M_{\widetilde{\chi}^0_1}$ plane are restored. The lesson of this exercise is that even in the popular and more general pMSSM model with 19 parameters more realistic bounds may be obtained if they are derived from data collected from a small number of channels. A similar observation was
made using LHC Run I data \cite{Chakraborti:2014gea,Chakraborti:2015mra}. One can, therefore,  go beyond rather contrived simplified models while analyzing LHC data without making the analysis involving many channels unmanageably complicated.

Various SUSY models with R-parity conservation\cite{drees2005theory,Baer:2006rs} provide a stable LSP. This sparticle, usually chosen as the lightest neutralino, is a popular DM candidate. This weakly interacting massive particle (WIMP) may explain the observed DM in the universe\cite{Lahanas:2003bh,Freedman:2003ys,Roszkowski:2004jc,Olive:2005qz,Baer:2008uu,Drees:2012ji,Arrenberg:2013rzp,Chattopadhyay:2006xb,Chattopadhyay:2007di,Godbole:2008it,Chattopadhyay:2008hk,Chattopadhyay:2009fr,Chattopadhyay:2010vp,Choudhury:2012kn,Mohanty:2012ri,Bhattacharya:2013uea,Harigaya:2014dwa,Bertone:2004pz,Jungman:1995df,Chakraborti:2017dpu,Delgado:2020url}. The first stringent constraint on the pMSSM came from the WMAP measurement \cite{Huang:2018xle} of the DM relic density. However, we shall use the more recent and slightly improved relic density data obtained by the PLANCK experiment \cite{Aghanim:2018eyx}.

In ref.\cite{Choudhury:2012tc} it was discussed that the physics of DM relic density depends on a multitudes of scenarios in the pMSSM. Apparently DM relic density production is expected to be almost insensitive to parameters in the strong sector. 
However, quite often a mass relation among the LSP-lighter chargino and gluino masses (like the one assumed by ATLAS mentioned above) is assumed to simplify the analysis. As a result the exclusion contours in the LSP-gluino mass plane from LHC data get further restricted by the DM data as will be shown in the upcoming sections. More interestingly gluino-LSP coannihilation, which has not been studied extensively \cite{Profumo:2004wk,Feldman:2009zc,Ellis:2015vaa,Nagata:2015hha,Ellis:2015vna}, directly plays an important role in DM relic density production in a wide class of models studied by us. This role of the gluino especially for high LSP mass has not been discussed in the context of recent LHC data.

We will further constrain the allowed parameter space (APS) of the above models using data from direct searches of DM from the LUX\cite{Akerib:2016vxi} and, subsequently, by the XENON1T\cite{Aprile:2018dbl} experiments. It should, however, be borne in mind that the computation of the LSP-nucleon scattering cross section, which is a crucial ingredient in this exercise, involves sizeable uncertainties both theoretical and experimental \cite{Ellis:2008hf,Ohki:2008ff,Giedt:2009mr,Perelstein:2012qg,Gondolo:2013xya}. For example, it is not known whether the flux of DM in the neighbourhood of the Earth is strong enough to lead to an observable signal. On the other hand the theory of LSP-nucleon scattering in a low energy experiment involves several uncertainties. Nevertheless, this terrestrial experiment for DM search remains popular. \\

In sec.\eqref{method} we shall briefly discuss the LHC and DM constraints and the strategy for simulating the LHC signals.  Descriptions of the models under study and the corresponding constraints from various data will be discussed in sec.\eqref{models}. Subsequently in sec.\eqref{SI} we will discuss the above models in the light of the direct detection of DM. We discuss the possibility of discriminating among several models, if SUSY signals show up in future LHC runs, in terms of relative signal strengths in sec.\eqref{rpm}. Finally, we will make concluding remarks in sec.\eqref{con}.
	
	\section{Methodology} \label{method}
	
	\subsection{Constraints} \label{cons}

	\subsubsection{Constraints from ATLAS $1l+jets+\met$ and $jets+\met$ analyses} \label{1lepton}
	
	In ref.\cite{Aad:2021zyy}, pair productions of gluinos and squarks at 13 TeV LHC with an integrated luminosity of $139$ \ifb were considered with subsequent decay of gluinos into $\widetilde{\chi}^\pm_1$ and appropriate quarks with $100\%$ BR. The lighter charginos further decay into $W^{\pm}$ and the LSP with $100\%$ BR leading to the signal comprising of $1l+jets+\cancel{E_T}$. The analysis was based on simplified model with certain assumptions mentioned in ref.\cite{Aad:2021zyy}. In this analysis ATLAS chose the mass hierarchy among $\widetilde{g}$, $\widetilde{\chi}^\pm_1$, $\widetilde{\chi}^0_1$ defined by $ x=\frac{M_{\widetilde{\chi}^\pm_1}-M_{\widetilde{\chi}^0_1}}{M_{\widetilde{g}}-M_{\widetilde{\chi}^0_1}}$, where $M_{\widetilde{g}}$, $M_{\widetilde{\chi}_1^\pm}$ and $M_{\widetilde{\chi}_1^0}$ are the masses of  gluino, lighter charginos and LSP respectively. This compression factor controls the mass differences between sparticles which means that increasing the gluino mass or the LSP mass results in enhancement of the chargino mass. Leptons considered here are only electrons or muons. The sets of cuts implemented were given in Table 2 in\cite{Aad:2021zyy} with respect to 4 different exclusive signal regions (SR) defined as SR2J, SR4J low-x, SR4J high-x, SR6J. Among these four SRs, only SR2J and SR6J are pertinent for $x=0.5$, i.e., the mass of $\widetilde{\chi}^\pm_1$ is placed midway between $M_{\widetilde{g}}$ and $M_{\widetilde{\chi}^0_1}$. Moreover, ATLAS collaboration also provided an exclusion contour as a function of $x$ for a fixed LSP mass. The upper limits on visible cross sections for different SRs for this study are adopted from Table 11 of ref.\cite{Aad:2021zyy}. From the exclusion contour in the $M_{\widetilde{g}} -M_{\lspone}$ plane shown in the top-left pane of Fig. 8 in ref.\cite{Aad:2021zyy}, it is observed that, for negligible LSP masses, $M_{\widetilde g}\lesssim 2.2 $ TeV is excluded. Moreover, for $\mlspone \gtrsim 1.26$ TeV there is no lower bound on mass of gluino.  \\
	
	Next we briefly discuss the gluino search by the ATLAS collaboration through gluino pair production leading to the final state comprising of $jets + E_T\!\!\!\!\!\!/~$~ at 13 TeV LHC with an integrated luminosity of $139$ \ifb. In simplified models only the LSP, the lighter charginos and the gluino are within the kinematic reach of the LHC whereas the other sparticles and heavy Higgses are set beyond the LHC reach. The mass of $\chonepm$ is considered as the arithmetic mean of the LSP and the gluino masses. In this analysis ATLAS assumed gluinos to be decaying into i$\rangle$ $\widetilde{g} \rightarrow q \bar{q}\widetilde{\chi}_1^0$ or in ii$\rangle$ $\widetilde{g} \rightarrow q \bar{q'}\widetilde{\chi}_1^\pm$ both with 100\% BR. These cascade decays finally lead to $jets+\met$ signature. We use the limits which are given for the direct decays of the gluino ($\widetilde{g} \rightarrow q \bar{q}\widetilde{\chi}_1^0$). Our analysis is based on this channel.
	
	 ATLAS collaboration consider ten inclusive SRs for studying this simplified model. The analysis cuts are given in Table 8 and Table 9 and the upper bound on effective cross sections of BSM physics for different SRs are given in Table 12 of ref.\cite{Aad:2020aze}. In Fig.14 of  ref.\cite{Aad:2020aze} it has been observed from the exclusion contour that, for negligible LSP mass $M_{\widetilde g}\lesssim 2.3$ TeV is excluded. Moreover, for $M_{\widetilde{\chi}_1^0} \gtrsim 1.2$ TeV there is practically no bound on $M_{\widetilde g}$.

	\subsubsection{Constraints from Dark Matter}
	
	Although the PLANCK data\cite{Aghanim:2018eyx} for satisfying the observed DM relic density has a tiny observational uncertainty ($0.120\pm0.001$), we note that there is about $10\%$ theoretical uncertainty in computing the SUSY DM relic density. This $10\%$ theoretical uncertainty arising from the renormalization scheme and the scale variations due to higher order SUSY-QCD corrections amounts to approximately six times the observational uncertainty as mentioned in refs.\cite{Harz:2016dql,Klasen:2016qyz}. Several (see for example, refs.\cite{Bertone:2015tza,Badziak:2017uto}) analyses included this theoretical error or even more in their works. This prompts us to consider $\Omega_\chi h^2=0.120\pm0.006$ that leads to the following bounds
	\begin{equation}
		0.114 < \Omega_\chi h^2 < 0.126     \nonumber
	\end{equation}
	\noindent  where $h=0.733\pm0.018$ \cite{Wong:2019kwg} is the expansion rate of the Universe, the Hubble constant, in the units of 100 km/Mpc-s. Apart from the above direct DM constraints we further impose the following bounds obtained from the indirect searches for the DM candidates.
	
	DM direct detection constraints on spin-independent (SI) LSP-proton scattering cross section $\sigma^{SI}_{\widetilde \chi p}$  is imposed using the the LUX\cite{Akerib:2016vxi} and the XENON1T\cite{Aprile:2018dbl} data.
	
	This analysis is also, in principle, subjected to the indirect detection bound on spin-dependent (SD) LSP-proton scattering cross section $\sigma^{SD}_{\widetilde \chi p}$ obtained from the IceCube experiment\cite{Aartsen:2016exj}. However, it is a well known fact that the IceCube constraints for a bino-like LSP are too weak and do not exclude any pMSSM points \cite{Silverwood:2012tp,Cahill-Rowley:2014boa}.

	\subsection{Simulation using LHC RUN-II data}
	\noindent We generate the sparticle mass spectra and decays in SUSY-HIT\cite{Djouadi:2006bz}. Gluino pairs are then produced using {\tt MG5 aMC@NLO}\cite{Alwall:2014hca} at 13 TeV centre of mass energy with the PDF set NNPDF2.3LO\cite{Ball:2012cx}. We have generated at least $5 \times 10^4$ signal events for various signal regions. 
	Next the showering and hadronization are processed via {\tt Pythia 8.2} \cite{Sjostrand:2014zea}. Matched events are generated using the CKKW-L scheme\cite{Lonnblad:2012ix}.
	{\tt Delphes 3.4}\cite{deFavereau:2013fsa} is employed for the reconstruction of dressed objects like jets, leptons, missing $p_T$ etc.
	and implementation of analysis cuts. The jets and leptons are preselected with certain loose quality requirements and are termed as `baseline' by the ATLAS collaboration. Signal objects, on the other hand, are further selected with tighter identification criteria applied on the baseline objects.
  Furthermore, the next to leading order (NLO) cross sections of the gluino pair production are computed by {\tt Prospino2.0} \cite{Plehn:2004rp}, for simulation purpose.
  
In  order  to  validate  our  simulation we compute the  exclusion  contour given by the ATLAS collaboration\cite{Aad:2021zyy,Aad:2020aze} using {\tt Delphes 3.4}\cite{deFavereau:2013fsa} after generating the events in {\tt MG5 aMC@NLO}\cite{Alwall:2014hca}. We have followed the electron-jets/muon-jets isolation criteria  for reconstructing and identifying those objects according to the ATLAS prescription. We have implemented the detector simulation using ATLAS card given in the {\tt Delphes 3.4} code after modifying the required pieces. The signal regions segmented by the ATLAS are characterized by a number of kinematical cuts and $N_{BSM}$ ($=$ production cross section $\times$ luminosity $\times$ cut efficiency $\times$ acceptance). 
Any pMSSM parameter space point is excluded in our simulation if the yield exceeds at least one of the values of $N_{BSM}$ obtained by the ATLAS collaboration. 
For the purpose of obtaining the exclusion contour for our models we have used the $N_{BSM}$ numbers corresponding to the SRs which provide the best sensitivity to the signal, as the limiting value. We vary $M_{\widetilde{\chi}_1^0}$ keeping $M_{\widetilde{g}}$ fixed in order to obtain the number of signal events in a particular model very close to $N_{BSM}$ value. The value of $M_{\widetilde{\chi}_1^0}$ for which the yield becomes closest to $N_{BSM}$ marks a limiting point for a particular $M_{\widetilde{g}}$ in a particular model. Next we vary the gluino mass in steps of 50 GeV to construct the entire exclusion contour in the $M_{\widetilde{g}}-M_{\widetilde{\chi}_1^0}$ plane. 
We have validated our simulation routine against the ATLAS analyses and have reproduced the relative efficiencies around $5\%$ accuracy.\footnote{{
 Cut flow tables for signal region SR2j (for $1l + jets + E_T\!\!\!\!\!\!/~$) and SR4j-3400 (for $jets+\met$) are given in the appendix \ref{appendix}. }}
The ATLAS exclusion contour reproduced by us is presented in all figures as the reference contour. The change in limits in different models are studied with respect to this reference contour.

	\section{Models} \label{models}	
	
This section has two subsections corresponding to different compositions of ${\widetilde{\chi}}^\pm_1 / {\widetilde{\chi}}^0_2$. These sparticles are assumed to be dominantly wino (higgsino) in subsec.\ref{wino}(\ref{higg}). In both the subsections the LSP is assumed to be mostly bino like. 
For wino type models the LSP is predominantly bino type over the parameter space we studied. However, in higgsino type models, there is, indeed, a significant amount of higgsino component in the LSP\cite{Chakraborti:2017dpu,Delgado:2020url} in the compressed region where the chargino and the LSP are close in terms of their masses. The heavier electroweakinos are assumed to be decoupled in both the discussions.
	
The masses of the LSP and gluino are free parameters and are determined by U(1) and SU(3) gaugino mass parameters $M_1$ and $M_3$, respectively. The SU(2) gaugino mass parameter $M_2$ is chosen in such a way to produce  $M_{\widetilde{\chi}^{\pm}_{1}}$ exactly halfway between $M_{\widetilde{g}}$ and $M_{\widetilde{\chi}^{0}_{1}}$. In the pMSSM scenario, $M_2$ sets the masses of both the lighter chargino  as well as the second lightest neutralino  which are nearly mass degenerate. All slepton mass parameters are set at $3.0$ TeV, and hence they are inaccessible at the LHC RUN II.

Third generation squarks are assumed to be decoupled. All other soft breaking trilinear terms are considered to be zero except the trilinear term for top quark. $A_t = 5$ TeV has been chosen in order to get SM-like Higgs mass at around $M_h \simeq 125$ GeV\cite{Degrassi:2002fi,Allanach:2004rh}. Throughout this study $\tan \beta$ is set at $10$ and the values of the pseudo scalar Higgs mass $M_A$ is chosen to be $3$ TeV and $M_{H^\pm}$ is obtained around 3.002 TeV for wino model (discussed in subsec.\ref{wino}) and for higgsino model (discussed in subsec.\ref{higg}). The choice of $M_A$ value is not particularly relevant for collider analysis, however, will become important in DM study. This particular value of $M_A$ is considered just for illustration without the loss of any generality.

	\subsection{The Wino Model} \label{wino}
	In wino model ${\widetilde{\chi}}^\pm_1 / {\widetilde{\chi}}^0_2$ is considered to be wino dominated. In this model we further assume ${\widetilde{\chi}}^0_1$ to be bino dominated in most of the parameter space. This can be achieved by considering a large value of higgsino mass parameter, $\mu$ ($\mu \simeq 2M_2$), which satisfies the hierarchy $M_1 < M_2 \ll \mu$, where $M_1$ and $M_2$ are the U(1) and SU(2) electroweak  gaugino mass parameters respectively. The BRs of three body decay modes of gluino for $M_{\widetilde{g}}=2.2$ TeV is shown in Fig.\eqref{comparison} against the ratio of first two generations of $L$ and $R$-squark masses. In addition, we have considered $M_{\widetilde{\chi}_1^\pm}= 1.1$ TeV and ${M_{\widetilde{\chi}^0_1}} \simeq 0$ for this plot. It can be easily inferred from the figure that for smaller ratio on the $x$-axis, there are large BRs for the $\widetilde{g}\rightarrow q \bar{q'} \widetilde{\chi}_1^\pm$ and $\widetilde{g}\rightarrow q \bar{q} \widetilde{\chi}_2^0$ modes. However, for larger values of that ratio, BR of $\widetilde{g}\rightarrow q \bar{q} \widetilde{\chi}_1^0$ mode becomes the dominant one.
	
		\begin{figure}[!h]
		\centering
		\includegraphics[scale=0.70]{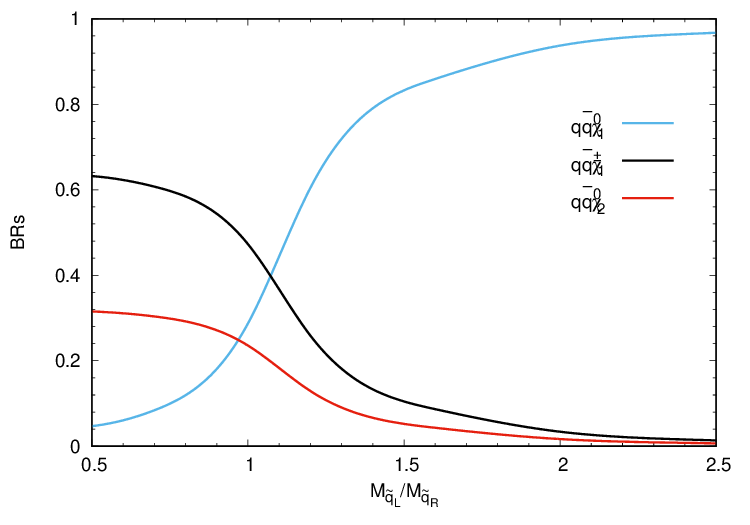}
		
		\caption{The BRs of three body decay modes of gluino is shown against the ratio of left and right squark masses of first two generations. The masses of the relevant particles are taken to be $M_{\widetilde{g}}=2.2$ TeV, $M_{\widetilde{\chi}_1^\pm}= 1.1$ TeV and ${M_{\widetilde{\chi}^0_1}} \simeq 0$. When the ratio of squark masses is small there are healthy BRs for $\widetilde{g}\rightarrow q \bar{q'} \widetilde{\chi}_1^\pm$ and $\widetilde{g}\rightarrow q \bar{q} \widetilde{\chi}_2^0$ channels and for the larger values of the corresponding ratio, the BR of $\widetilde{g}\rightarrow q \bar{q} \widetilde{\chi}_1^0$ mode becomes the dominant one.}
		\label{comparison}
	\end{figure}

	In this analysis, $M_2$ is not a free parameter rather it is constrained by the relation $x=\frac{M_{\widetilde{\chi}^\pm_1}-M_{\widetilde{\chi}^0_1}}{M_{\widetilde{g}}-M_{\widetilde{\chi}^0_1}}$ (see subsec.\eqref{1lepton}). Here, increasing the value of either $M_1$ or $M_3$ increases the value of $M_2$. Since $\mu$ is proportional to $M_2$, therefore,  the higgsino content in lighter chargino decreases, which is well-suited for our purpose to generate wino-type lighter chargino, at par with the ATLAS analysis. On top of it, we have a number of variant scenarios assuming different hierarchy of masses of the $L$-type and $R$-type squarks, which we describe in the following subsections.

	\subsubsection{LLRLW: Left Light Right Light Wino model }\label{lrew}
	
	In the Left Light Right Light Wino (LLRLW) model, it is assumed that the diagonal entries of the first two generations of squark mass matrices 
		take the value around $2.5$ TeV, which are heavier than the  gluino masses we study. As a result gluino has only three body decay modes and in particular, it has moderately large BR in the decay mode $\widetilde{g} \rightarrow q \bar{q} \widetilde{\chi}_1^0$. This is because of the fact that the intermediate off-shell squarks have equally large doublet and singlet contributions, where the latter favourably couples to the bino like LSP. Therefore, the BR ($\widetilde{g} \rightarrow q q' \widetilde{\chi}_1^\pm$) is depleted compared to the simplified version considered by the ATLAS which assumed this BR to be  $100\%$.  As a result, the yield of $1l + jets + E_T\!\!\!\!\!\!/~$ signal events which diminishes appreciably 
		the LSP mass compared to that of the ATLAS simplified scenario corresponding to the same gluino mass.

	In Fig.\eqref{llrl_fig1} the parameter space excluded by the ATLAS collaboration is displayed by the solid black line and the corresponding degraded exclusion contour in respect of LLRLW model is shown by the green dashed line. From the ATLAS analysis \cite{Aad:2021zyy} it has been observed that for negligible LSP mass $M_{\widetilde{g}} < 2.2$ TeV is excluded. Furthermore, there is no bound on $M_{\widetilde{g}}$ where $M_{\widetilde{\chi}_1^0}$ exceeds 1.26 TeV. It is evident from the figure that the limits on gluino mass reduces appreciably by an amount around ($350-450$) GeV with respect to the corresponding ATLAS bound. For negligible LSP mass the excluded region is now relaxed to  $M_{\widetilde{g}}\approx 1.82$ TeV, whereas, for $M_{\widetilde{\chi}_1^0}\approx 750$ GeV the bound on $M_{\widetilde{g}}$ ceases to exist.
	
 
 In Table\eqref{table2}, we showcase the limits on gluino mass for different compression factors $x$ as defined in previously in sec.\eqref{1lepton}. For the purpose of illustration, we obtain the limits using  $1l+jets+\met$ and $jets+\met$ data for two different LSP masses ($M_{\widetilde{\chi}_1^0} \simeq 0$ and $M_{\widetilde{\chi}_1^0} = 500$ GeV) for the LLRLW model. It is observed that in the $1l + jets + \met$ search channel, changes in the compression factor have a larger impact on the bound compared to the other channel.

	\FloatBarrier

\begin{table}[h!]	
\begin{center}
		
\begin{tabular}{||c|c|c|c||}
\hline
Compression factor ($x$) & $M_{\widetilde{\chi}_1^0}$ (GeV) & \makecell{Search channel \\(ATLAS) } & Limit on $M_{\widetilde{g}}$ (GeV) \\
\hline

\multirow{4}{*}{0.4} & \multirow{2}{*}{$\simeq 0$} & 1$\ell$ & 1860 \\
\cline{3-4}
 & & 0$\ell$ & 2110 \\
\cline{2-4}
& \multirow{2}{*}{$500$} & 1$\ell$ & 1800 \\
\cline{3-4}
& & 0$\ell$ & 2005 \\
\hline

\multirow{4}{*}{0.5} & \multirow{2}{*}{$\simeq 0$} & 1$\ell$ & 1815 \\
\cline{3-4}
& & 0$\ell$ & 2135 \\
\cline{2-4}
& \multirow{2}{*}{$500$} & 1$\ell$ & 1780 \\
\cline{3-4}
& & 0$\ell$ & 2010 \\
\hline

\multirow{4}{*}{0.6} & \multirow{2}{*}{$\simeq 0$} & 1$\ell$ & 1730 \\
\cline{3-4}
& & 0$\ell$ & 2150 \\
\cline{2-4}
& \multirow{2}{*}{$500$} & 1$\ell$ & 1270 \\
\cline{3-4}
& & 0$\ell$ & 2020 \\
\hline

\hline
\multirow{4}{*}{0.7} & \multirow{2}{*}{$\simeq 0$} & 1$\ell$ & 1435 \\
\cline{3-4}
& & 0$\ell$ & 2160 \\
\cline{2-4}
& \multirow{2}{*}{$500$} & 1$\ell$ & - \\
\cline{3-4}
& & 0$\ell$ & 2025 \\
\hline
	
\end{tabular} 
\end{center}
	
\caption{ Limits on the gluino mass for different values of compression factor ($x$) for LLRLW model in the light of $1l+jets+\met$ (shown as 1$\ell$) and $jets+\met$ (shown as 0$\ell$) data obtained by the ATLAS collaboration. '-' denotes that there is no bound on the gluino mass for this LSP mass. }
\label{table2}      		
\end{table}

	\begin{figure}[!h]
		\centering
		\includegraphics[scale=0.75]{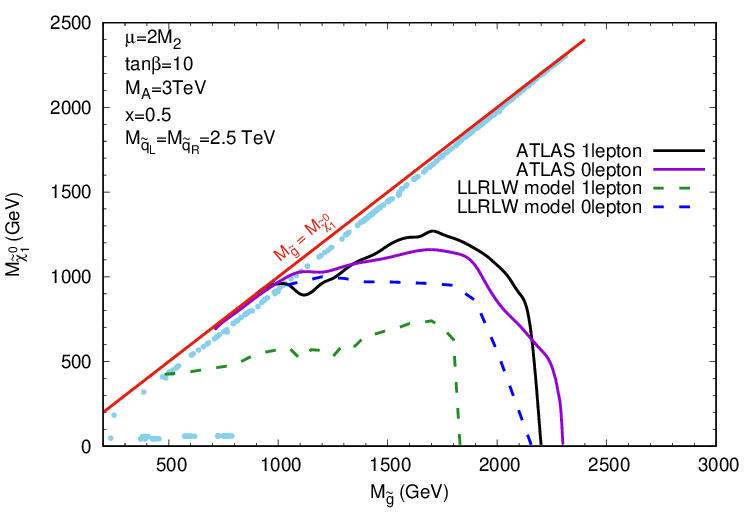}
		
		\caption{Exclusion contour for the Left Light Right Light Wino (LLRLW) model is shown in the $M_{\widetilde{g}}- M_{\widetilde{\chi}_1^0}$ plane. The solid black line and solid violet line correspond to the observed limits as given by ATLAS from the $1l + jets + \cancel{E_T}$ data\cite{Aad:2021zyy} and $jets+\met$ data\cite{Aad:2020aze} respectively at 13 TeV center-of-mass energy in the simplified model. The dashed green curve is the modified excluded region for the LLRLW model for the final state comprising of $1l + jets + \cancel{E_T}$ where relatively low $L$-type and $R$-type squark masses (but still heavier than the gluino mass) are considered along with wino type lighter charginos and second lightest neutralino. The exclusion contour for this model for the final state $jets+\met$ is also shown by dashed blue curve. The cyan blue dots signify masses of the LSP which are allowed by the PLANCK relic density data\cite{Aghanim:2018eyx}. The region above the red diagonal line is disallowed as gluino becomes the LSP. }
		\label{llrl_fig1}
	\end{figure}

	The two distinct cyan blue dotted branches in the Fig.\eqref{llrl_fig1} are in agreement with the PLANCK data\cite{Aghanim:2018eyx}. 
    The corresponding allowed regions are shown in cyan blue points. 
    In the lower branch the cyan blue points produce the correct value of the DM relic density corresponding to LSP-pair annihilation through a s-channel light h-resonance. Moreover, there are points in that branch which correspond to $Z$-resonances also. This branch extends up to $M_{\widetilde{g}} \lesssim 800$ GeV. For higher gluino masses the tiny higgsino component of the dominantly bino like LSP is too small to continue this mechanism. 
	
	The upper branch is approximately parallel to the red line. The lower part of this branch represents the mass region where the LSP pair annihilation is the dominant DM relic density production mechanism. In the upper part of this branch the LSP coannihilation with a nearly mass degenerate sparticle like the $\widetilde{g}$, $\widetilde{\chi}_1^\pm$ and $\widetilde{\chi}_2^0$ are the main contribution to the relic density production. For very high gluino masses beyond the LHC reach, LSP coannihilation with gluino plays a very important role for DM relic density production. 
	
	\subsubsection{LLRHW: Left Light Right Heavy Wino model \label{llrhw}}
This model is closer to the ATLAS simplified model compared to the LLRLW model. In the Left Light Right Heavy Wino (LLRHW) model, the SU(2) doublet $L$-type squark mass parameters of first two generations continue to be $\sim 2.5$ TeV as in the previous case. However, the masses of the $R$-type squarks are considered to be at large value around $6.5$ TeV. This means that the $R$-type squarks play insignificant role in gluino decay processes for this scenario. Since the intermediate off-shell squarks are mostly SU(2) doublet, the BR of the decay mode, $\widetilde{g} \rightarrow q q'{\widetilde{\chi}}^\pm_1$ is higher than that of the LLRLW model. However, in this scenario gluino also decays to ${\widetilde{\chi}}^{0}_{2} q \bar{q}$ with a moderate to large BR. ${\widetilde{\chi}}^{0}_{2}$ being wino like, further, predominantly decays to $h \widetilde{\chi}_1^0$ along with $Z \widetilde{\chi}_1^0$ decay mode with negligible BR. The decay mode $h \widetilde{\chi}_1^0$ or $Z \widetilde{\chi}_1^0$ does not contribute to the $1l + jets + E_T\!\!\!\!\!\!/~$ final state. Consequently, the number of $1l + jets + E_T\!\!\!\!\!\!/~$ signal events decreases appreciably compared to the simplified model considered by the ATLAS collaboration. In this model, the exclusion contour shrinks by a modest amount $\simeq (300-350)$ GeV in comparison to the exclusion contour of the ATLAS experiment. For negligible LSP mass the exclusion region degrades to  $M_{\widetilde{g}}\approx 1.9$ TeV, whereas, for $M_{\widetilde{\chi}_1^0}\approx 920$ GeV the bound on $M_{\widetilde{g}}$ goes away. As expected the exclusion limits in the LLRHW model gets depleted by a smaller amount compared to the previous LLRLW model. 


 In Table\eqref{table3} we show the variation of the limits on the gluino mass for different compression factors for $1l+jets+\met$ and $jets+\met$ final states for two different values of LSP masses in LLRHW model.

	\FloatBarrier

\begin{table}[h!]	
\begin{center}
		
\begin{tabular}{||c|c|c|c||}
	
\hline
Compression factor ($x$) & $M_{\widetilde{\chi}_1^0}$ (GeV) & \makecell{Search channel \\(ATLAS)} & Limit on $M_{\widetilde{g}}$ (GeV) \\
\hline

\multirow{4}{*}{0.4} & \multirow{2}{*}{$\simeq 0$} & 1$\ell$ & 1940 \\
\cline{3-4}
& & 0$\ell$ & 2110 \\
\cline{2-4}
& \multirow{2}{*}{$500$} & 1$\ell$ & 1850 \\
\cline{3-4}
& & 0$\ell$ & 2000 \\
\hline

\multirow{4}{*}{0.5} & \multirow{2}{*}{$\simeq 0$} & 1$\ell$ & 1912 \\
\cline{3-4}
& & 0$\ell$ & 2125 \\
\cline{2-4}
& \multirow{2}{*}{$500$} & 1$\ell$ & 1822 \\
\cline{3-4}
& & 0$\ell$ & 2005 \\
\hline

\multirow{4}{*}{0.6} & \multirow{2}{*}{$\simeq 0$} & 1$\ell$ & 1890 \\
\cline{3-4}
& & 0$\ell$ & 2130 \\
\cline{2-4}
& \multirow{2}{*}{$500$} & 1$\ell$ & 1750 \\
\cline{3-4}
& & 0$\ell$ & 2010 \\
\hline

\multirow{4}{*}{0.7} & \multirow{2}{*}{$\simeq 0$} & 1$\ell$ & 1795 \\
\cline{3-4}
& & 0$\ell$ & 2140 \\
\cline{2-4}
& \multirow{2}{*}{$500$} & 1$\ell$ & 1540 \\
\cline{3-4}
& & 0$\ell$ & 2020 \\
\hline
			
	\end{tabular} 
	\end{center}
	
	\caption{ Limits on the gluino mass for different values of compression factor ($x$) for LLRHW model in the light of $1l+jets+\met$ and $jets+\met$ data obtained by the ATLAS collaboration. Conventions are same as Table.\eqref{table2}.}
	\label{table3}      		
\end{table}

In Fig.\eqref{llrh_fig2} we present the PLANCK data\cite{Aghanim:2018eyx} satisfying points over the LHC exclusion plot similar to that discussed in subsec.\eqref{lrew} for Fig.\eqref{llrl_fig1}.

\begin{figure}[!htb]
	\centering
	
	\centering
	\includegraphics[scale=0.75]{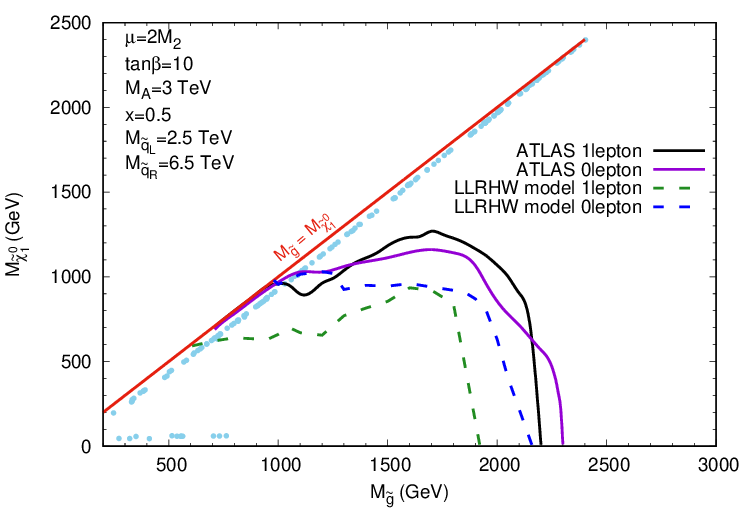}
	
	\caption{Exclusion contour (dashed green line and dashed blue line) is obtained for the Left Light Right Heavy Wino (LLRHW) model where  $L$-type squark masses are small, and $R$-type squark masses are quite heavy ($M_{\widetilde{g}} < M_{\widetilde{q}_L} \ll M_{\widetilde{q}_R}$). The lighter charginos and second lightest neutralino are wino type. The ATLAS exclusion limits on gluino masses in their simplified model is displayed by solid black line. The cyan points satisfy relic density data obtained from the PLANCK data\cite{Aghanim:2018eyx}. The colour conventions are the same as given in Fig.\eqref{llrl_fig1}.}
	\label{llrh_fig2}

\end{figure}

	In another variant of wino model namely, Left Heavy Right Light Wino (LHRLW) model, this $1l+jets+\met$ signal disappears as the gluino predominantly decays into $jets+\met$ channel.

	\subsection{Higgsino model} \label{higg}
	
	We now discuss the scenario where the lighter chargino (${\widetilde{\chi}}^\pm_1 $), the second and third lightest neutralinos (${\widetilde{\chi}}^0_2$, ${\widetilde{\chi}}^0_3$) are higgsino dominated and ${\widetilde{\chi}}^0_1$ being still bino dominated. This model can be realized by considering large $M_2$. The preferred hierarchy of the gaugino and higgsino mass parameters are to be set as $M_1 < \mu \ll M_2 $. In the following subsections we will study this model for various scenarios depending on where $L$-type or $R$-type squark masses are placed with respect to the gluino mass.

	\subsubsection{LLRLH: Left Light Right Light Higgsino model} \label{llrlh}
	In the Left Light Right Light Higgsino (LLRLH) model, it is assumed that both the $L$-type and $R$-type squark mass parameters of the first two generations are fixed at $\simeq 2.5$ TeV.  
	For this part of the parameter space, the gluino pair production dominantly gives  $jets + E_T\!\!\!\!\!\!/~$~ signal. The BR of gluino decaying into $\widetilde{\chi}^\pm_1$/$\widetilde{\chi}^0_2$ in association with quarks is almost negligible, resulting in yield of a meagre amount of $1l+jets+\cancel{E_T}$ final states. As a result, the bound on $M_{\widetilde{g}}$ from $1l+jets+\cancel{E_T}$\cite{Aad:2021zyy} search by ATLAS almost vanishes for higgsino model.

%
%
%
%
%

	\subsubsection{LLRHH: Left Light Right Heavy Higgsino model \label{llrhh}}
	
	In the Left Light Right Heavy Higgsino (LLRHH) model, the mass of the $L$-type squarks of first two generations are assumed to have values around $2.5$ TeV, whereas, the $R$-type squarks are set at $ 6.5$ TeV. Therefore, the $R$-type squarks have less chance to appear in the off-shell propagator in gluino decay. The dominant decay modes of the gluino for this case is $q\bar{q} \widetilde{\chi}^0_1$, resulting in very depleted number of $1l+jets+\cancel{E_T}$\cite{Aad:2021zyy} final states. For such scenario, therefore, the ATLAS bounds practically the same as in the previous case.

	We note in passing that for the case where the $R$-type squarks are made to be lighter and $L$-type squarks are heavier, the model is equivalent to the ATLAS simplified model for search in the $jets+\cancel{E_T}$\cite{Aad:2020aze} final state which we will discuss in the next subsection.
	
	\subsubsection{Higgsino models in the light of $jets+\cancel{E_T}$ data} \label{0lepton}
	
	As noted in subsecs.\eqref{llrlh} and \eqref{llrhh}, in all the variants of higgsino type models, practically all the parameter space are allowed by $1l+jets+\met$ data\cite{Aad:2021zyy}. This happens because, the gluino dominantly decays into the LSP and jets in such models making the one lepton signal vanishingly small. Therefore, the viability of such models need to be  analyzed again by $jets+\met$ data\cite{Aad:2020aze}.

	\begin{figure}[!htb]
		\centering
		
		\centering
		\includegraphics[scale=0.75]{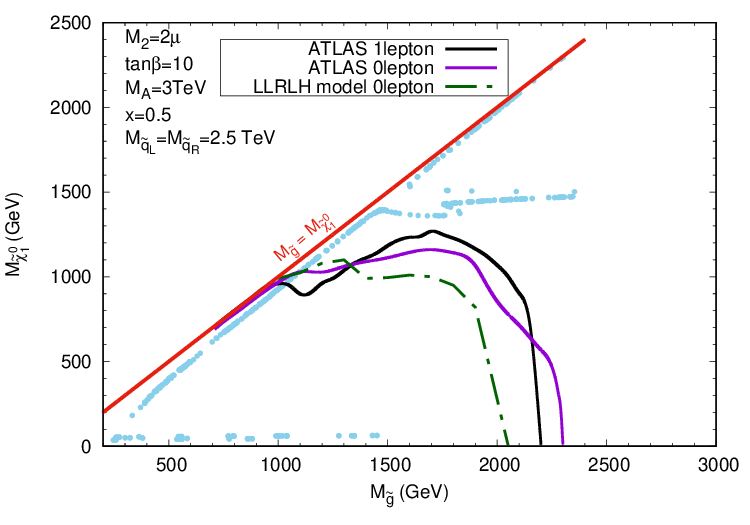}
		
		\caption{Exclusion contour (dashed green line) is obtained for the Left Light Right Light Higgsino (LLRLH) model where both $L$-type and $R$-type squark masses are small but larger than the gluino mass ($M_{\widetilde{g}} < M_{\widetilde{q}_L} = M_{\widetilde{q}_R}$). The lighter charginos and second lightest neutralino are higgsino type. The ATLAS exclusion limits on gluino mass for the search of $1l+jets + \cancel{E_T}$ and $jets + \cancel{E_T}$ \cite{Aad:2021zyy,Aad:2020aze} at 13 TeV for simplified model are displayed in solid black line and solid violet line respectively. The colour conventions are the same as given in Fig.\eqref{llrl_fig1}.}
		
		\label{fig_10}

	\end{figure}

For the purpose of illustration we consider the LLRLH model which is unbounded from the ATLAS  $1l+jets+\met$\cite{Aad:2021zyy} constraint. In Fig.\eqref{fig_10} we present the exclusion contour for ATLAS $jets+\met$ data in respect of simplified model by solid violet line. In the same figure the green dashed line corresponds to exclusion contour for LLRLH model. It is also seen from the figure that the gluino mass bounds do not alter appreciably for that obtained from the $1l+jets+\met$ analysis for negligible LSP mass. The disallowed region of parameter space (near around $M_{\widetilde{g}}=M_{\widetilde{\chi}^{0}_{1}}$ line) extends a little above the ATLAS $jets+\met$ exclusion contour thereby making the bound on $M_{\widetilde{g}}$ stronger. 
 It should be kept in mind that the whole region of parameter space is allowed by $1l+jets+\met$ data.

We also present the PLANCK data\cite{Aghanim:2018eyx} allowed points for the LLRLH model in Fig.\eqref{fig_10}. In the figure we observe three separate branches. The lowest one produces the right relic abundance through the LSP-pair annihilation via $s$-channel $h/Z$ resonances. 
The diagonal branch, almost overlapping with the degenerate gluino-LSP line, is interesting and is of particular importance. The lower half of this branch (till about $M_{\widetilde{g}} \sim 700$ GeV), though ruled out by collider data, gives correct relic density due to large pair annihilation of LSP to $t\bar{t}$, $W^+ W^-$, $ZZ$ through s-channel Higgs facilitated by the enhanced higgsino component in LSP . Though, annihilation to $W^+ W^-$ and $ZZ$ also receive contributions from the exchange of t-channel lightest chargino and  and second lightest neutralino. The coannihilation with the $\widetilde{\chi}_1^{\pm}/\widetilde{\chi}_2^{0}$ also contributes, but with smaller fraction. The upper half of this diagonal branch corresponds to the gluino coannihilation channel, as discussed in last paragraph of subsec.~\ref{lrew} . 

The nearly horizontal branch at $M_{\widetilde{\chi}_1^{0}} \sim 1500$ GeV appears due to the resonant LSP pair annihilation for $M_A = M_{H^{\pm}} = 3.0$ TeV. For any $M_A$ value, such branch would appear at $M_{\widetilde{\chi}_1^{0}} \simeq \frac{1}{2}M_A$. However, we have considered this value of $M_A$ just for illustration as this evades $M_A-\tan \beta$ bound.
	
 We observe from Fig.\eqref{llrl_fig1} (Fig.\eqref{llrh_fig2}) that for negligible LSP mass the limit on $M_{\widetilde{g}}$ is obtained around 1.82 TeV (1.9 TeV) for LLRLW (LLRHW) models, constrained by $1l+jets+\met$ data. Compared to these numbers, in LLRLH model the bound on $M_{\widetilde{g}}$ becomes 2.1 TeV obtained from $jets+\met$ data. Thus the bounds on $M_{\widetilde{g}}$ in LLRLH model (see Fig.\eqref{fig_10}) is quite comparable to that for wino type LLRLW/LLRHW model (see Fig.\eqref{llrl_fig1}/\eqref{llrh_fig2}).

	\section{Direct Detection through Spin-independent scattering } \label{SI}

	In addition to the constraint from the PLANCK results, we compute the spin-independent (SI) LSP-proton scattering cross section ($\sigma^{SI}_{\chi p}$) in relation to the earlier LUX data \cite{Akerib:2016vxi} and the more recent XENON1T data \cite{Aprile:2018dbl}. The cross section is sensitive to the mass of the LSP and the compositions of the LSP\cite{Jungman:1995df,Drees:1993bu}. We must keep in mind various sizeable uncertainties which plague the computation of $\sigma^{SI}_{\chi p}$. For example, they may arise due to low energy hadronic physics \cite{Ellis:2008hf,Ohki:2008ff,Giedt:2009mr,Perelstein:2012qg}, the local DM density which has not been measured directly  \cite{Beskidt:2012bh,Beskidt:2012sk,Bovy:2012tw}, non-Maxwellian velocity distribution of the WIMPs \cite{Bhattacharjee:2012xm,Fairbairn:2012zs} etc. Combining all these  the rate of direct detection  may involve an order of magnitude of uncertainty or more.

	Within the detector nucleus, SI interaction occurs between the lightest neutralino (LSP) and the quarks inside the nucleon through $s$-channel squark exchange and $t$-channel Higgs exchange processes. The dominant contribution to $\sigma^{SI}_{\chi p}$  comes from the $t$-channel Higgs exchange as squarks are assumed to be considerably heavy throughout this analysis. The effective couplings are dependent on the nature of composition of the LSP. The products of the gaugino and the higgsino components of the neutralino diagonalizing matrix have a major contribution in the $h (H) \widetilde{\chi}^0_1 \widetilde{\chi}^0_1$ coupling where $H$ is the CP even heavy neutral Higgs boson. Assuming the LSP to be bino dominated $(M_1 \ll M_2,\mu)$, the effective couplings can be approximated as \cite{Hisano:2009xv}, 
	\begin{multicols}{2}
		\begin{eqnarray}	
			\noindent	C_{h \widetilde{\chi} \widetilde{\chi}} \simeq \frac{M_Z s_W t_W}{M_1^2-\mu^2}[M_1+\mu \sin 2\beta] ~~~ \nonumber
		\end{eqnarray} 
		
		\begin{eqnarray}
			C_{H \widetilde{\chi} \widetilde{\chi}} \simeq -\frac{M_Z s_W t_W}{M_1^2-\mu^2}\mu \cos 2\beta  
			\nonumber
		\end{eqnarray}
	\end{multicols}
	\noindent where, $s_W= \sin \theta_W$ with $\theta_W$ as the Weinberg angle and $M_Z$ is the mass of the $Z$ boson. From the couplings we can easily interpret that a significant amount of bino-higgsino mixing i.e. $M_1 \simeq \mu$ results in a large value of the SI cross section $\sigma^{SI}_{\chi p}$, whereas, a bino dominated LSP has small coupling and hence the  XENON1T\cite{Aprile:2018dbl} experiment can probe such models. In the following subsections the computed results are presented for different wino and higgsino models.

	\subsection{ LLRLW and LLRHW } \label{winod}
	We now consider the points in Figs.\eqref{llrl_fig1} and \eqref{llrh_fig2} which are consistent with both the LHC and the PLANCK data and test their viability vis-a-vis the XENON1T \cite{Aprile:2018dbl} data.

\begin{figure}[h!]
		\centering
	\begin{subfigure}{.5\textwidth}
			\centering
	\includegraphics[width=.9\linewidth]{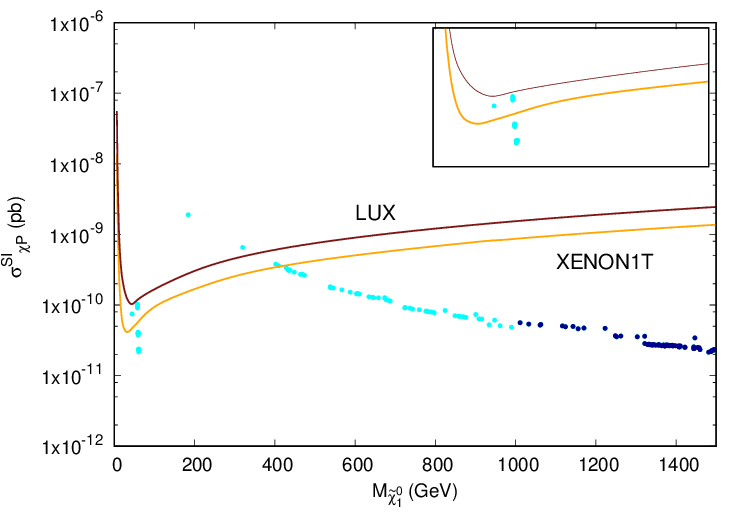}
			\caption{}
			\label{figlre}
		\end{subfigure}%
		\begin{subfigure}{.5\textwidth}
			\centering
	\includegraphics[width=.9\linewidth]{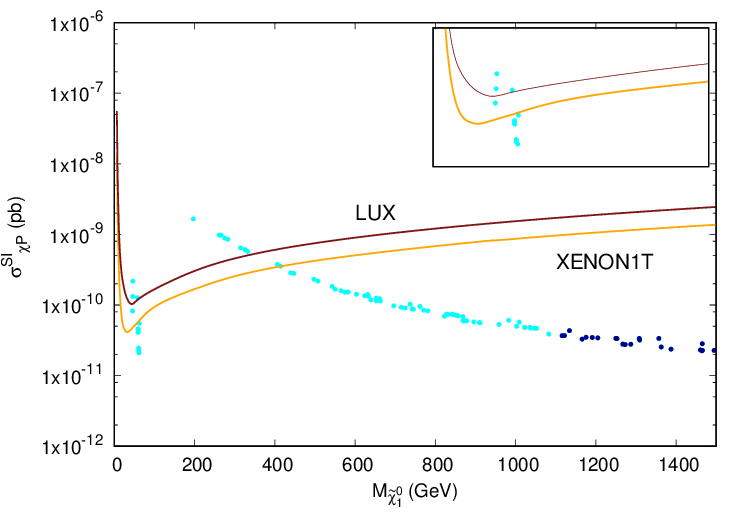}
			\caption{}
			\label{figll}
		\end{subfigure}
	
		\caption{Spin-Independent direct detection cross section for the a) Left Light Right Light Wino (LLRLW) model and b) Left Light Right Heavy Wino (LLRHW) model. Upper limits on the spin-independent elastic WIMP-nucleon cross section at $90\%$ C.L provided by the LUX\cite{Akerib:2016vxi} and the XENON1T\cite{Aprile:2018dbl} are shown as brown and orange lines respectively. The cyan points only satisfy the relic density data from the PLANCK\cite{Aghanim:2018eyx} experiment. The points which are allowed by both the LHC and PLANCK data are shown by the dark blue points.}
		
		\label{figwinosi}
	\end{figure}

	In Fig.\eqref{figlre} we present the plot in the $M_{\widetilde{\chi}^0_1}-\sigma^{SI}_{\chi p}$ plane for LLRLW model. The value of  $\sigma^{SI}_{\chi p}$ is computed using {\tt micrOMEGAs 5.2}\cite{Belanger:2020gnr}. The exclusion limits for the LUX and the XENON1T results are shown by the brown and orange curves respectively. 
	 The cyan points denote the region of parameter space allowed by the PLANCK data\cite{Aghanim:2018eyx} only.  Whereas the dark blue region of parameter space consistent with both LHC\cite{Aad:2021zyy} and PLANCK data satisfies the latest direct detection constraints \cite{Akerib:2016vxi,Aprile:2018dbl}. The similar observations are found in LLRHW model which is displayed in Fig.\eqref{figll}.

	We also observe in Fig.\eqref{figlre} that the  points which satisfy the condition $M_{\widetilde{\chi}_1^0} \gtrsim 1.0$ TeV are allowed by the ATLAS data\cite{Aad:2021zyy,Aad:2020aze} (see Figs.\eqref{llrl_fig1}$-$\eqref{fig_10}), the PLANCK constraint\cite{Aghanim:2018eyx} and as well as the DM direct detection results\cite{Akerib:2016vxi,Aprile:2018dbl}. From Fig.\eqref{figll} for LLRHW model we further infer that the parameter space points with $M_{\widetilde{\chi}_1^0} \gtrsim 1.1$ TeV are also allowed by all the constraints discussed above.\\

	\subsection{Higgsino models }
	
		\begin{figure}[h!]
		\centering
		\includegraphics[scale=0.75]{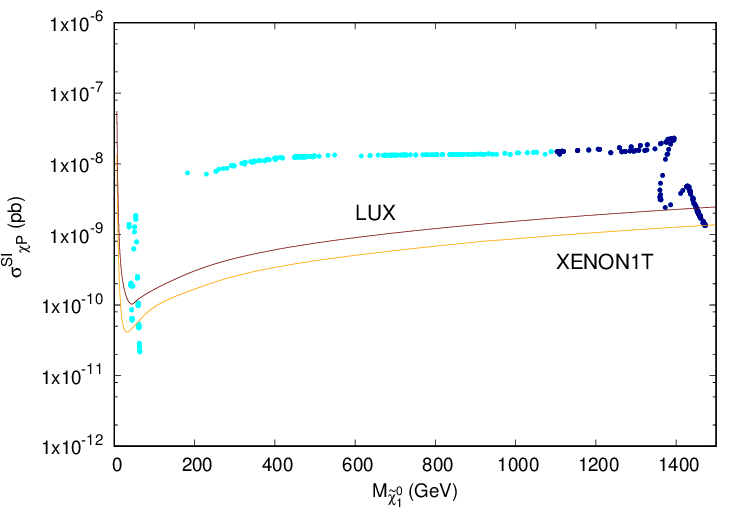}
		\caption{ Spin-Independent direct detection results for the Left Light Right Light Higgsino (LLRLH) model. Colours and conventions are the same as in Fig.\eqref{figwinosi}}
		\label{figlrehiggsi}
	\end{figure}
	
	For the sake of completeness, we plot the $M_{\widetilde{\chi}_1^0}-\sigma^{SI}_{\chi p}$ for the LLRLH model in Fig.\eqref{figlrehiggsi}. It appears that the points consistent with the LHC and PLANCK data are in conflict with XENON1T data. 
The theoretical uncertainties in DM direct detection include local DM density, Maxwellian velocity distribution of WIMPs and the hadronic uncertainties arising primarily from strangeness content of the nucleons inside the colliding nuclei. The estimates of the uncertainties for local DM density and the Maxwellian velocity distribution of WIMPs can result in about 10\% of their respective central value \cite{Green:2017odb}. Previously the hadronic uncertainties would have changed the base value even by an order of magnitude. However, in view of recent lattice QCD calculations this uncertainty reduces to about $10\%$ (see for e.g., \cite{Ellis:2018dmb,DelNobile:2021wmp}).

 Therefore, some of the excluded points in Fig.\eqref{figlrehiggsi} by direct detection experiments may become allowed due to these fluctuations and vice versa. Due to the stringency of this constraint most of the parameter space for higgsino type scenarios allowed by PLANCK data can not be experimentally probed.

	\section{Relative signal strengths of various models} \label{rpm}

	From our analyses in two different gluino search channels, we sketch the feasibility of distinguishing different pMSSM scenarios discussed in sec.\eqref{models}. If gluino signals are observed in at least two channels, this method might be helpful to identify the underlying model. In order to keep the production cross section fixed, we choose $M_{\widetilde{g}}$ = 1.3 TeV (allowed by all data), while the other mass parameters are chosen from various models. To carry out our analysis we choose three benchmark points (BPs) from Fig.\eqref{llrl_fig1}$-$\eqref{fig_10} (see Table\eqref{bpspec}). BP1, BP2 and BP3 are chosen from LLRLW, LLRHW and LLRLH models respectively.
	
    The decay modes and their branching fractions relevant for the gluino signals for $M_{\widetilde{g}}$ = 1.3 TeV are presented in Table\eqref{bpdecay}. It may be noted that in this table the BRs of the gluino may not add up to 100\%. This is due to the fact that in some scenarios the gluino also decays into the other channels with small but non-negligible BRs. However, all modes are taken into account while simulating the gluino signal.

	      		\FloatBarrier
	
	\begin{table}[h!]	
		\resizebox{\textwidth}{!}{
			\begin{tabular}{||c|c|c|c|c|c|c|c|c|c|c||}
				\hline \hline
				Benchmark point & $M_1$ & $M_2$ & $M_3$ & $\mu$ &$M_{\widetilde{g}}$ &$M_{\widetilde{\chi}_1^0}$ & $M_{\widetilde{\chi}_1^\pm}$ & $\Omega_{\widetilde{\chi}} h^2$  & $\sigma_{SI} $ (pb) & $\sigma_{SD}$ (pb) \\
				
				\hline

				BP1 (From Fig.\eqref{llrl_fig1}) & $1269$ & $1236$ & $982$ & $2472$ & $1300$ & $1238$ & $1269$ & $0.116$ & $3.052 \times 10^{-11}$ & $3.702 \times 10^{-10}$ \\
				
				\hline
				
				BP2 (From Fig.\eqref{llrh_fig2}) & $1262$ & $1232$ & $946$ & $2464$ & $1300$ & $1231$ & $1265$ & $0.124$ & $2.844 \times 10^{-11}$ & $7.571 \times 10^{-9}$ \\
				
				\hline

				BP3 (Form Fig.\eqref{fig_10}) & $1310$ & $2492$ & $983$ & $1246$ & $1300$ & $1238$ & $1268$ & $0.122$ & $1.737 \times 10^{-8}$ & $6.022 \times 10^{-6}$ \\

				\hline \hline

		\end{tabular} }

		\caption{ The sparticle mass spectra corresponding to different benchmark points chosen from Fig.2 to Fig.4. The mass parameters are in GeV.}
		
			\label{bpspec}
		
	\end{table}

	\FloatBarrier
	
	\begin{table}[h!]	
		\begin{center}

			\begin{tabular}{||c|c|c|c||}
				\hline  \hline
				
				Decay Modes & BP1 & BP2 & BP3  \\
				\hline
				
				$\widetilde{g} \rightarrow \widetilde{\chi}_1^0 q \bar{q}$ & $80.86$ & $28.08$ & $64.34$  \\

				$ \rightarrow \widetilde{\chi}_2^0 q \bar{q}$ & $6.16$ & $23.20$ & $-$  \\

				$ \rightarrow \widetilde{\chi}_1^\pm q \bar{q}'$ & $12.68$ & $47.60$ & $-$  \\
				
				$ \rightarrow \widetilde{\chi}_1^0 g$ & $-$ & $-$ & $29.68$  \\
				
				$ \rightarrow \widetilde{\chi}_2^0 g$ & $-$ & $-$ & $5.78$ \\
			\hline \hline
		$\widetilde{\chi}_1^\pm \rightarrow \widetilde{\chi}_1^0 W^{*\pm}  $ & $100$ & $100$ & $100$  \\
				       				    				    \hline \hline
				       				
		$ \widetilde{\chi}_2^0  \rightarrow\widetilde{\chi}_1^0 q \bar{q} $ & $82.14$ & $84.13$ & $67.56$  \\

		$\rightarrow \widetilde{\chi}_1^0 l^+ l^- $ & $3.52$ & $3.67$ & $32.07$ \\		
		
	$\rightarrow \widetilde{\chi}_1^0 \gamma $ & $12.67$ & $10.49$ & $0.33$ \\		       			
				
				\hline \hline

			\end{tabular} 
		\end{center}

		\caption{	The BR (\%) of the dominant decay modes of the sparticles for the different BPs. Here $q$ and $q'$ denote first two generation of quarks . `$-$' denotes a negligible branching fraction and $W^{*\pm}$ denotes the off-shell $W^\pm$.}	
		
		\label{bpdecay}
		
	\end{table}
	
	\FloatBarrier
	
	\begin{table}[h!]	
		\begin{center}
			
			\begin{tabular}{||c|c|c|c||}
				\hline  \hline			
				
				Points & $S_1$ & $S_0$ & $R$ \\
				\hline
				BP1 & $1.28$ & $10.48$ & 0.78 \\
				\hline
				BP2 & $5.05$ & $10.65$ & 0.35 \\
				\hline
				BP3 & 0.27 & 15.67 & 0.97 \\
				
				\hline \hline
				
			\end{tabular} 
		\end{center}
		
		\caption{ The table displays  $S_0$ and $S_1$ and $R$ for different BPs. $S_0$ and $S_1$ represent the number of signal events corresponding to $jets+\met$ data  and $1l+jets+\met$ data respectively for an integrated luminosity 139 fb$^{-1}$. Here $R$ stands for $\frac{S_0 - S_1}{S_0 + S_1}$. }
		\label{tablebp}      		
		\end{table}
	
	We now use the ATLAS gluino searches for $\sqrt{s} = 13$ TeV with integrated luminosity $\mathcal{L} = 139$ fb$^{-1}$ in $1l + jets + \cancel{E_T}$ and $jets + \cancel{E_T}$ channels, as described in subsec.\eqref{1lepton} and in sec.\eqref{0lepton}. We define an observable called relative signal strength, $R = \frac{S_0 - S_1}{S_0 + S_1}$, where $S_0$ and $S_1$ denote the number of signal events of $jets + \cancel{E_T}$ and $1l + jets + \cancel{E_T}$ channels respectively. It is worth mentioning that the ratio $R$ is almost free from theoretical uncertainties like the choice of QCD scale and the PDFs etc.

	In Table\eqref{tablebp} we display the magnitudes of $S_1$, $S_0$ and $R$ corresponding to different BPs. We find that $S_0$ is larger than $S_1$ in all models. However, the ratio $R$, distinguishes different models. This technique of discriminating among various pMSSM models using multichannel analysis may be a useful tool if significant signals in different channels are found in future LHC experiments. A similar method was used to illustrate this possibility of discrimination using RUN-I data\cite{Chakraborti:2014gea,Chatterjee:2014uda,Chakraborti:2015mra}.

	\section{Conclusion} \label{con}
	
	The main purpose of this paper is to re-examine the limits on $M_{\widetilde{g}}$ obtained by the ATLAS collaboration using rather contrived simplified models \cite{Aad:2021zyy,Aad:2020aze}. We analyze the same data in the more generic pMSSM scenario and study the variation of the limits with parameters which are most relevant for the  signal being analyzed. We also present the variation of limits with the compression factors for both the wino models.
	
	We first revisit the ATLAS $1l+jets+\met$ data \cite{Aad:2021zyy}. Here we assume LSP to be bino like and $\widetilde{\chi}_1^\pm$, $\widetilde{\chi}_2^0$ to be wino like (see subsec.\eqref{wino}). Other electroweakinos are assumed to be decoupled. In the susbsec.\eqref{lrew} both $L$-type and $R$-type squarks are assumed to be light and mass degenerate ($M_{\widetilde{q}_L}=M_{\widetilde{q}_R}\approx2.5$ TeV). The resulting exclusion contour is displayed in the $M_{\widetilde{g}}-M_{\widetilde{\chi}_1^0}$ plane in Fig.\eqref{llrl_fig1}. The substantial change in the exclusion contour with respect to the ATLAS contour (due to reasons discussed in the text) can be seen in Fig.\eqref{llrl_fig1}. However, if the compression factor $x$ (defined in the subsec.\eqref{1lepton}) is altered compared to $x=0.5$, the corresponding contour gets modified accordingly. For example if $x=0.7$, the limit on $M_{\widetilde{g}}$ is around 1.43 TeV for negligible LSP mass.
	
 On the other hand if $M_{\widetilde{q}_R}\gg M_{\widetilde{q}_L} \approx2.5$ TeV, the pMSSM framework is closer to the ATLAS simplified model. This is reflected by the fact that the new exclusion contour shrinks modestly compared to the ATLAS one (see Fig.\eqref{llrh_fig2}).
	
	The results change dramatically, if $\widetilde{\chi}_1^\pm$, $\widetilde{\chi}_2^0$ and $\widetilde{\chi}_3^0$ are higgsino like (see subsec.\eqref{higg}). In this scenario the gluino dominantly decays directly into the LSP and jets. Thus irrespective of the squark mass hierarchy the $1l+jets+\met$ signature is highly depleted. 
	
	We also analyze both the wino and higgsino models in the light of ATLAS $jets+\met$ data. For wino models (LLRLW and LLRHW) the corresponding exclusion contours using $jets+\met$ data are shown in figs.\eqref{llrl_fig1} and \eqref{llrh_fig2}. In fig.\eqref{fig_10} we present a comparative study of the constraints in the LLRLH model and that with ATLAS simplified model (both $1l+jets+\met$ and $jets+\met$). 
	 There is no significant change observed between the two. From Figs.\eqref{llrl_fig1}$-$\eqref{fig_10} we find that even in the compressed scenario ($M_{\widetilde{g}} \simeq M_{\widetilde{\chi}_1^0}$), $M_{\widetilde{g}}\gtrsim1.5$ TeV is a fairly conservative limit\cite{Feng:1999zg,Feng:2013pwa}. For smaller LSP mass $M_{\widetilde{g}}\gtrsim2$ TeV is a fairly reliable bound. More importantly it asserts the well-known fact that the bounds from a single channel in a simplified model is not reliable. This we have seen in the case of higgsino models where the constraint from the $1l+jets+\met$ ceases to exist, but $jets+\met$ puts bound in place (see the higgsino models in subsec.\eqref{higg}). Therefore, a multi-channel analysis may lead to reliable limits even in the more generic pMSSM model with 19 parameters.
	
	As discussed in the introduction one can make a useful analysis of the interplay between the LHC and PLANCK data on dark matter relic density within the framework of the pMSSM. From Figs.\eqref{llrl_fig1}$-$\eqref{fig_10} 
	we find that coannihilation of the LSP with other spin half gauginos is an important channel which produces correct value of the DM relic density in the parameter space allowed by the LHC data. In particular the gluino-LSP coannihilation plays a very crucial role among the coannihilation processes.  
	For the higgsino models there is another interesting process for DM relic density production. They are LSP annihilation via $H^+$ or $A$ resonance (see Fig.\eqref{fig_10}).
	
	In Fig.\eqref{figlre}(\eqref{figll}) we show the points in the LLRLW (LLRHW) model consistent with both  LHC and PLANCK data by the dark blue points. All presently available constraints are satisfied by these points for $M_{\widetilde{\chi}_1^0}\gtrsim1.1$ TeV (see Fig.\eqref{fig_10}).
	
	We have displayed the direct detection results in connection with XENON1T data for LLRLH model in Fig.\eqref{figlrehiggsi}. The points satisfying LHC and PLANCK constraints are found to be apparently in conflict  with XENON1T data. This disagreement, however, is not conclusive if theoretical and experimental uncertainties in deriving the limits are taken into consideration.


Finally, we have carried out a small exercise in sec.\eqref{rpm}. This  illustrates that if future LHC experiments discover signals in
both $jets+\met$ and $1l+jets+\met$ channels then the underlying  wino and higgsino type models can be
distinguished from each other by measuring an observable, $R$ defined in sec.\eqref{rpm}.

\appendix   

\section{Comparing relative cut efficiencies} \label{appendix}
Cut-flow tables for signal region SR2j (for $1l + jets + E_T\!\!\!\!\!\!/~$) and SR4j-3400 (for $jets+\met$) are displayed for illustration. The relative cut efficiencies are shown for comparing the agreement between  the corresponding ATLAS analysis and that of our MC simulation.\\

\subsection{Comparing cut efficiencies for $1l+jets+\met$ channel}

\begin{table}[htb]
    \centering
    \resizebox{0.8\textwidth}{!}{
    \begin{tabular}{|c|c|c|}
        \hline
        Selection cuts & \makecell[c]{Relative efficiency\\ in $\%$\\by the ATLAS collaboration} & \makecell[c]{Relative efficiency\\ in $\%$\\by our analysis}\\
        \hline
        Preselction cut & 100  &  100 \\
        \hline
        $p_{T}^{lep} < 25 $ GeV  & 21.1  & 23.4 \\
        \hline
        $E_T^{miss} > 400$ GeV & 35.1  &  43.4\\
        \hline
        $N_{b-jet} = 0$ &  75.9 &  81.1\\
        \hline
        $E_T^{miss}/m_{eff} > 0.25$ & 99.0 & 95.1 \\
        \hline
        $N_{jet} > p_{T}^{lep} /10$ &  100 &  100  \\
        \hline
        $m_{eff} > 700$ GeV  &   100 &  100 \\
        \hline
        $m_{T} > 100 $ GeV &  75 &  75.5 \\
        \hline
         $ 700$ GeV $< m_{eff} < 1300$ GeV &    40.0  &  36.5\\
        \hline
    \end{tabular}}
    \caption{Comparison between relative cut-flow ($C_{i+1}/C_i$) corresponding to ATLAS data\cite{Aad:2021zyy} and our simulation for the signal region SR2j. The numbers correspond to $M_{\widetilde{g}}$=1000 GeV and $M_{\widetilde{\chi}_1^0}$=800 GeV for the $1l+jets+\met$ final state.}
   \label{tab:my_label2}
\end{table}

\clearpage

\subsection{Comparing cut efficiencies for $jets+\met$ channel}

\begin{table}[htb]
    \centering
    \resizebox{0.8\textwidth}{!}{
    \begin{tabular}{|c|c|c|}
        \hline
        Selection cuts & \makecell[c]{Relative efficiency\\ in $\%$\\ by the ATLAS collaboration} & \makecell[c]{Relative efficiency\\ in $\%$ \\ by our analysis}\\
        \hline
        Preselction and cleaning cut & 100 & 100 \\
        \hline
        Jet multiplicity $ \ge $ 4 & 93.9 &  96.0 \\
        \hline
        $\Delta\phi(j_{1,2,(3)},E_T^{miss})_{min} > 0.4$ & 83.7 &  83.1\\
        \hline
        $\Delta\phi(j_{i>3},E_T^{miss})_{min} > 0.2$ & 88.8 & 85.4\\
        \hline
        $p_{T}(j_4) > 100$ GeV  &  88.7 &  91.4\\
        \hline
        $|\eta(j_{1,2,3,4})| < 2.0$   & 91.0 &  91.3 \\
        \hline
        Aplanarity $ > 0.04 $ &  69.26 & 70.7\\
        \hline
        $E_T^{miss}/\sqrt{H_T} > 10$ GeV$^{1/2}$ &  87.6 &  88.2\\
        \hline
        $m_{eff} > 3400$ GeV &  41.0 &  40.7\\
        \hline
        
    \end{tabular}}
    \caption{Comparison between relative cut-flow ($C_{i+1}/C_i$) corresponding to ATLAS data\cite{Aad:2020aze} and our simulation for the signal region SR4j-3400. The numbers correspond to $M_{\widetilde{g}}$=2200 GeV and $M_{\widetilde{\chi}_1^0}$=600 GeV for the $jets+\met$ final state.}
    \label{tab:my_label}
\end{table}

\clearpage

	\section*{Acknowledgements}
	
	Authors are grateful to Prof. Amitava Datta for his guidance and valuable suggestions on the work and careful reading of the manuscript. Authors acknowledge the help of Dr. Nabanita Ganguly in computation. AM is also thankful to Dr. Jyoti Prasad Saha and Md. Raju for discussions. AM would like to thank DST for providing the INSPIRE Fellowship [IF170693].

	\bibliographystyle{hephys}
	\bibliography{references}

\end{document}